\batchmode
\makeatletter
\def\input@path{{/home/flatmax/flatmax/personal.work/research/fxt/mattAndHarvey/}}
\makeatother
\documentclass[australian,rmp,altaffilletter]{article}
\usepackage[T1]{fontenc}
\usepackage[latin9]{inputenc}
\usepackage{amsmath}
\usepackage{amssymb}

\makeatletter

\providecommand{\tabularnewline}{\\}

\usepackage{endnotes}


\makeatother

\usepackage{babel}
\begin{document}
\title{Scaling the time and Fourier domains to align periodically and their
convolution}
\author{Matt Flax, W. Harvey Holmes}
\maketitle
\begin{abstract}
This note shows how to align a periodic signal with its the Fourier
transform by means of frequency or time scaling. This may be useful
in developing new algorithms, for example pitch estimation. This note
also convolves the signals and the frequency time convolution is denoted
``fxt''.\footnote{This is a recasting of Flax's original version \cite{Flax2016}. The
following also includes some additional explanatory and other material.
However, the effects of noise or uncertain parameters haven't yet
been investigated - however intuition tell us that noise which is
localised in the time domain will not be localised in the Fourier
domain and vice versa which is an advantage to overcoming certain
types of noise.}
\end{abstract}

\section{Introduction}

Suppose $x\left(t\right)$ is a periodic signal with period $t_{p}$
and fundamental frequency 
\begin{equation}
f_{p}=1/t_{p}\label{eq:fundamental frequency}
\end{equation}
In the frequency domain there will be sinusoidal components at integer
multiples of the fundamental $f_{p}$; i.e. at 
\begin{equation}
f=n\,f_{p},\,\,n\in\mathbb{Z}\label{eq:harmonic frequencies}
\end{equation}

These correspond to impulses in the Fourier transform $X\left(f\right)$,
defined as

\begin{equation}
{\displaystyle X(f)}\triangleq\int_{-\infty}^{\infty}x\left(t\right)e^{-j2\pi ft}dt
\end{equation}

For this signal we can state the following:
\begin{quote}
\textit{}%
\begin{minipage}[t]{1\columnwidth}%
\begin{quote}
\textit{Important fact}: $X\left(f\right)$ has non-zero frequency
domain samples only at multiples of the frequency $f_{p}$.

\textit{General case}: The non-zero samples at multiples of $f_{p}$
will generally not be of equal amplitude.

\textit{Special case}: However, if $x\left(t\right)$ consists of
a train of identical impulses at multiples of $t_{p}$, then $X\left(f\right)$
also consists of a train of identical impulses spaced at multiples
of $f_{p}$.
\end{quote}
\end{minipage}
\end{quote}
The general case of periodic $x\left(t\right)$ can be reduced to
the special case as follows. Suppose the shape of a single period
of $x\left(t\right)$ is 
\begin{equation}
h\left(t\right)=\left\{ \begin{array}{cc}
x\left(t\right), & t\in\left[-t_{p}/2,\,t_{p}/2\right]\\
0, & t\notin\left[-t_{p}/2,\,t_{p}/2\right]
\end{array}\right.
\end{equation}
and that a train of unit impulses is 
\begin{equation}
s\left(t\right)=\sum_{n=-\infty}^{\infty}\delta\left(t+nt_{p}\right)
\end{equation}
with\footnote{\texttt{http://www.dsprelated.com/freebooks/sasp/Impulse\_Trains.html}}
\begin{equation}
S\left(f\right)=f_{p}\sum_{m=-\infty}^{\infty}\delta\left(f+mf_{p}\right)
\end{equation}
This is also a train of equal impulses in the frequency domain. This
gives us a convolutional representation of $x\left(t\right)$: 
\begin{equation}
x\left(t\right)=s\left(t\right)\ast h\left(t\right)\label{eq:convolution}
\end{equation}
with 
\[
X\left(f\right)=S\left(f\right)H\left(f\right)
\]
Hence the special case above applies if we compare $x\left(t\right)$
with $S\left(f\right)$ instead of $X\left(f\right)$. That is,
\begin{quote}
\textit{}%
\begin{minipage}[t]{1\columnwidth}%
\begin{quote}
\textit{General case}: If the periodic signal $x\left(t\right)$ is
decomposed as in (\ref{eq:convolution}) with $s\left(t\right)$ consisting
of a train of identical impulses at multiples of $t_{p}$, then $S\left(f\right)$
is also periodic and consists of a train of identical impulses spaced
at multiples of $f_{p}$. However, $X\left(f\right)$ still consists
of a train of impulses spaced at multiples of $f_{p}$, but these
impulses are not in general identical.
\end{quote}
\end{minipage}
\end{quote}
This establishes a sort of duality between the time and frequency
domains that is only valid for periodic signals. This fact is exploited
in this paper to scale $S\left(f\right)$ so that it is aligned with
$x\left(t\right)$. Alternatively, a dual scaling can be used to scale
$s\left(t\right)$ so that it is aligned with $X\left(f\right)$. 

As a result it is hoped that we can possibly extract extra information
from the signal by comparing the two domains in new ways, for example
to enhance pitch estimation.

\section{Theory in the sampled finite length case}

In the following we consider only finite length sampled data signals,
so that $X\left(f\right)$ will be the discrete Fourier transform
(DFT \cite{oppenheim99}) of $x\left(t\right)$. Suppose $N$ samples
are taken at a sampling rate $f_{s}$ (Hz). These samples are therefore
spaced apart at 
\begin{equation}
\delta_{t}\triangleq f_{s}^{-1}\,\mathrm{(seconds)}
\end{equation}
and the total duration of the signal is 
\begin{equation}
T=\left(N-1\right)\delta_{t}\,\mathrm{(seconds)}
\end{equation}
If we write
\begin{equation}
x_{n}\triangleq x\left(n\delta_{t}\right),\,\,n=0:N-1
\end{equation}
where $N$ is the total number of samples, then the DFT is defined
as the $N$-vector $X$ with the elements
\begin{eqnarray*}
X_{k} & \triangleq & \sum_{n=0}^{N-1}x_{n}e^{-j2\pi kn/N},\,\,k=0:N-1
\end{eqnarray*}

\begin{quote}
\textit{}%
\begin{minipage}[t]{1\columnwidth}%
\begin{quote}
\textit{Note on frequency scaling}: The index $k$ represents frequency
in the following way\footnote{There are probably better ways of showing this.}.
If $x$ is the complex sinusoid $x\left(t\right)=e^{j2\pi ft}$ at
frequency $f$, then $x_{n}=e^{j2\pi fn\delta_{t}}=e^{j2\pi nf/f_{s}}$
and
\begin{eqnarray*}
X_{k} & = & \sum_{n=0}^{N-1}e^{j2\pi fn\delta_{t}}e^{-j2\pi kn/N}\\
 & = & \sum_{n=0}^{N-1}e^{j2\pi n\left(f/f_{s}-k/N\right)ft}
\end{eqnarray*}
The amplitude $\left|X_{k}\right|$ is maximized at\footnote{Rounding is needed because $X_{k}$ is only defined for integer values
of $k$.}
\[
k\approx\mathrm{round}\left(N\dfrac{f}{f_{s}}\right)\equiv\mathrm{round}\left(\dfrac{f}{\delta_{f}}\right)
\]
which implies that $k$ may be considered to be a scaled frequency
representation. Note that $k_{\max}=N-1$ corresponds to $f=\dfrac{N-1}{N}f_{s}\approx f_{s}$.
Also, each increment of $k$ corresponds to a frequency increment
of $\delta_{f}=f_{s}/N$. These facts help in scaling Matlab frequency
plots. See \texttt{testdft.m}.
\end{quote}
\end{minipage}
\end{quote}
In the rest of this note we ignore the fact that the period $t_{p}$
may not really be an exact multiple of $\delta_{t}$, and assume that
it is (however, it may be desirable in Matlab to enforce this condition).
Then the number of samples in a fundamental period is
\begin{eqnarray}
N_{t} & = & \dfrac{t_{p}}{\delta_{t}}\nonumber \\
 & = & t_{p}f_{s}
\end{eqnarray}
where $N_{t}\in\mathbb{Z}$. The only non-zero DFT frequency components
are spaced at $f_{p}$ Hz, given by (\ref{eq:fundamental frequency}). 
\begin{quote}
\begin{minipage}[t]{1\columnwidth}%
\begin{quote}
\texttt{\textit{Note}}: Because of the sampling, we must assume that
the maximum frequency present is $f_{s}/2$ (the Nyquist frequency)
- i.e. we make the assumption that $f_{s}$ is large enough for there
to be no aliasing. That is, from (\ref{eq:harmonic frequencies})
$nf_{p}\leq f_{s}/2,\,\,n\in\mathbb{Z}$, so that the maximum number
of harmonics that we can consider in the sample is 
\begin{equation}
n_{\mathrm{max}}=\left\lfloor \dfrac{1}{2}\dfrac{f_{s}}{f_{p}}\right\rfloor 
\end{equation}
\end{quote}
\end{minipage}
\end{quote}
However, the DFT of the sampled signal will have $N$ Fourier components
in the full range $f\in\left[0,\,f_{s}\right]$. These are spaced
at intervals of 
\begin{equation}
\delta_{f}=\dfrac{f_{s}}{N}\,\mathrm{(Hz)}\label{eq:dft spacing}
\end{equation}
(Unless $f_{s}$ is an exact multiple of $f_{p}$, or equivalently
that the period $t_{p}$ is an exact multiple of $\delta_{t}$, none
of these Fourier samples will exactly coincide with the actual harmonic
frequencies $nf_{p},\,\,n\in\mathbb{Z}$.)

In the DFT, the harmonics are spaced at intervals of $N_{f}$ samples,
with
\begin{eqnarray}
N_{f} & = & \dfrac{f_{p}}{\delta_{f}}\nonumber \\
 & = & N\frac{f_{p}}{f_{s}}
\end{eqnarray}

\subsection{Key variables}
\begin{quote}
\begin{minipage}[t]{1\columnwidth}%
\begin{quote}
\begin{tabular}{ll}
$t_{p}=1/f_{p}$  &
Signal period (s)\tabularnewline
$f_{p}=1/t_{p}$  &
Signal frequency (Hz)\tabularnewline
$f_{s}$  &
Sampling frequency (Hz)\tabularnewline
$\delta_{t}=1/f_{s}$ &
Sample interval (s)\tabularnewline
$\delta_{f}=fs/N$ &
Spacing of DFT frequency points (Hz)\tabularnewline
$N$ &
Number of samples in signal (and in its DFT)\tabularnewline
$N_{t}=t_{p}/\delta_{t}=t_{p}f_{s}$ &
Number of time samples in a signal period\tabularnewline
$N_{f}=f_{p}/\delta_{f}$ &
Spacing of harmonics in the DFT (samples)\tabularnewline
\end{tabular}
\end{quote}
\end{minipage}
\end{quote}

\subsection{Resampling of the frequency domain signal}

We wish to resample to equalize the number of samples between major
components in the time and frequency domains. There are two cases,
depending on which signal ($X\left(f\right)$ or $x\left(t\right)$)
is resampled. In each case we wish to have the same total number $N$
of samples after resampling, so that they can be compared.

First, we will resample the DFT signal $X\left(f\right)$ so that
the spacing $N_{f}$ of the harmonics in $X\left(f\right)$ is the
same as the number of samples $N_{t}$ in a period of $x\left(t\right)$.
That is, we will change $N_{f}$ to $N'_{f}\triangleq aN_{f}$ such
that $aN_{f}=N_{t}$. Hence the scale factor required is
\begin{equation}
a=\dfrac{N_{t}}{N_{f}}
\end{equation}

The total number of frequency samples would then be $aN$ instead
of $N$. To retain the same total number of samples, this means that
the frequency increments must change from $\delta_{f}$ to $\delta'_{f}\triangleq\dfrac{\delta_{f}}{a}$.
Hence the new frequency increment is
\begin{eqnarray}
\delta'_{f} & = & N_{f}\dfrac{1}{N_{t}}\delta_{f}\nonumber \\
 & = & \frac{f_{p}}{\delta_{f}}\dfrac{\delta_{t}}{t_{p}}\delta_{f}\nonumber \\
 & = & f_{p}^{2}\delta_{t}\nonumber \\
 & = & \dfrac{f_{p}^{2}}{f_{s}}\\
 & \equiv & \dfrac{1}{t_{p}^{2}f_{s}}
\end{eqnarray}

It follows that the range of frequencies in the resampled DFT (still
of length $N$) will change from $\left[0,\,f_{s}\right]$ to 
\begin{equation}
\left[0,\,\left(N-1\right)\dfrac{f_{p}^{2}}{f_{s}}\right]
\end{equation}

\subsubsection{Interpolation of $X\left(f\right)$}

It will be necessary to interpolate the DFT to produce $N$ values
over the above frequency range. Knowing the index $n_{\mathrm{end}}$
(in the vector $X$) of the new end frequency 

\begin{equation}
f_{\mathrm{end}}=\left(N-1\right)\dfrac{f_{p}^{2}}{f_{s}}
\end{equation}
is useful when doing the interpolation using \texttt{interp1.m} in
Matlab. Allowing for the fact that Matlab indices start from 1 instead
of 0, this index is given by $\dfrac{n_{\mathrm{end}}-1}{N}=\dfrac{f_{\mathrm{end}}}{f_{s}}$;
i.e. 
\begin{eqnarray}
n_{\mathrm{end}} & = & \dfrac{Nf_{\mathrm{end}}}{f_{s}}+1\nonumber \\
 & = & N\left(N-1\right)\dfrac{f_{p}^{2}}{f_{s}^{2}}+1
\end{eqnarray}

This is the same as Flax's formula \cite{Flax2016}
\begin{eqnarray*}
\left(N-1\right)Mf+1 & = & \left(N-1\right)\dfrac{N}{f_{s}^{2}t_{p}^{2}}+1\\
 & = & N\left(N-1\right)\dfrac{f_{p}^{2}}{f_{s}^{2}}+1
\end{eqnarray*}

in Matt's code (\texttt{fxtEx21.m}).

\subsection{Resampling of the time domain signal}

In this case we will resample $x\left(t\right)$ so that the number
of samples in a period $N_{t}$ is the same as the number $N_{f}$
of samples between harmonics in $X\left(f\right)$. That is, we will
change $N_{t}$ to $N'_{t}\triangleq bN_{t}$ such that $bN_{t}=N_{f}$.
Hence the scale factor required ismy
\begin{equation}
b=\dfrac{N_{f}}{N_{t}}\equiv1/a
\end{equation}

The total number of time samples will then be $bN$ instead of $N$.
To retain the same total number of samples, this means that the time
increments must change from $\delta_{t}$ to $\delta'_{t}\triangleq\dfrac{\delta_{t}}{b}$.
Hence the new time increment is
\begin{eqnarray}
\delta'_{t} & = & N_{t}\dfrac{1}{N_{f}}\delta_{t}\nonumber \\
 & = & \frac{t_{p}}{\delta_{t}}\dfrac{\delta_{f}}{f_{p}}\delta_{t}\nonumber \\
 & = & t_{p}^{2}\delta_{f}\nonumber \\
 & = & \dfrac{t_{p}^{2}f_{s}}{N}\\
 & \equiv & \dfrac{f_{s}}{Nf_{p}^{2}}
\end{eqnarray}

It follows that the time range in the resampled signal (still of length
$N$) will change from $\left[0,\,\left(N-1\right)\delta_{t}\right]$
to 
\begin{equation}
\left[0,\,\left(N-1\right)\dfrac{t_{p}^{2}f_{s}}{N}\right]
\end{equation}

\subsubsection{Interpolation of $x\left(t\right)$}

It will be necessary to interpolate $x\left(t\right)$ to produce
$N$ values over the above time range. Knowing the index $m_{\mathrm{end}}$
(in the vector $x$) of the new final time 
\begin{equation}
t_{\mathrm{end}}=\left(N-1\right)\dfrac{t_{p}^{2}f_{s}}{N}
\end{equation}
is useful when doing the interpolation using \texttt{interp1.m} in
Matlab. Allowing for the fact that Matlab indices start from 1 instead
of 0, this index is given by $\dfrac{m_{\mathrm{end}}-1}{N}=\dfrac{t_{\mathrm{end}}}{N\delta_{t}}$;
i.e. 
\begin{eqnarray}
m_{\mathrm{end}} & = & \dfrac{t_{\mathrm{end}}}{\delta_{t}}+1\nonumber \\
 & = & \left(N-1\right)\dfrac{t_{p}^{2}f_{s}^{2}}{N}+1\\
 & \equiv & \dfrac{N-1}{N}\dfrac{f_{s}^{2}}{f_{p}^{2}}+1
\end{eqnarray}

This is the same as Flax's formula \cite{Flax2016} 
\begin{eqnarray*}
\left(N-1\right)Mt+1 & = & \left(N-1\right)\dfrac{f_{s}^{2}t_{p}^{2}}{N}+1
\end{eqnarray*}

in Flax's code (\texttt{fxtEx21.m}).

\subsection{Flax's original version \cite{Flax2016} (using this new notation)}

We wish to rescale $f_{p}$ so that there are an equivalent number
of samples between both the time period and the Fourier harmonics,
call the frequency scaling coefficient $a$, then
\begin{eqnarray}
t_{p} & = & af_{p}\label{eq:new domain equation}\\
N_{t}\delta_{t} & = & aN_{f}\delta_{f}\nonumber \\
t_{p}f_{s}\delta_{t} & = & af_{p}\frac{N}{f_{s}}\delta_{f}\nonumber \\
af_{p} & = & t_{p}\frac{f_{s}^{2}}{N}\frac{\delta_{t}}{\delta_{f}}\label{eq:scaled equivalence}
\end{eqnarray}
or
\[
a=\frac{t_{p}^{2}f_{s}^{2}}{N}\frac{\delta_{t}}{\delta_{f}}
\]

In the classical signal processing $a=1$ and there is a well known
inverse relationship between time period and harmonic distance ($t_{p}=f_{p}^{-1}$),
which when combined with Equation \ref{eq:scaled equivalence} yields
a constrained relationship between time and frequency sample duration/distance
respectively which is 
\begin{eqnarray}
t_{p} & = & \frac{1}{f_{p}}\label{eq:classical inverse equation}\\
t_{p} & =\frac{\delta_{f}}{\delta_{t}} & \frac{N}{f_{s}^{2}t_{p}}\nonumber \\
\frac{\delta_{t}}{\delta_{f}} & = & \frac{N}{f_{s}^{2}t_{p}^{2}}\label{eq:df scaling}
\end{eqnarray}

\section{Conclusion}

Prior to this article the only commonly known equivalence between
time duration and Fourier distance was the inverse relation (\ref{eq:fundamental frequency})
between period and frequency for a periodic signal. For sampled data
signals this article goes further.

Using the above theory, it is now possible to resample the signals
in either the frequency or time domains so that the sample count between
Fourier harmonics in the frequency domain is the same as the number
of samples in a period in the the time domain. 

The possible uses of this theory (for example pitch detection) have
still to be explored. An important issue is that the above theory
assumes the period $t_{p}$ is known, which means that in practice
this parameter must often first be estimated.

Similarly, the effect of noise or inexact $t_{p}$ has to be evaluated,
as we will rarely have an uncontaminated periodic signal with exactly
known $t_{p}$.

The same approach defined in this article can be used to derive scaling
coefficients for any other linear transformation.

\bibliographystyle{plain}
\bibliography{0_home_flatmax_flatmax_personal_work_research_fxt_mattAndHarvey_bib}

\end{document}